\documentstyle[amsfonts,aps,prd,epsf,floats]{revtex}

\newcommand{\beq}{\begin{equation}}
\newcommand{\eeq}{\end{equation}}
\newcommand{\beqa}{\begin{eqnarray}}
\newcommand{\eeqa}{\end{eqnarray}}
\newcommand{\non}{\nonumber}

\begin{document}

\draft

\title{
\hfill {\rm OUTP-98-60-P} \\
Instantons and the QCD Vacuum Wavefunctional}

\author{
William E. Brown\thanks{E-mail: w.brown1@physics.ox.ac.uk.
Address after 15 September 1998: Theoretical Physics, The Rockefeller
University, 1230 York Avenue - New York, NY 10021},
Juan P. Garrahan\thanks{E-mail: j.garrahan1@physics.ox.ac.uk.
On leave from Departamento de F\'{\i}sica, Universidad de Buenos Aires,
Ciudad Universitaria, 1428 Buenos Aires, Argentina.},
Ian I. Kogan\thanks{E-mail: i.kogan1@physics.ox.ac.uk. 
On leave from ITEP, B. Cheremyshkinskaya 25, Moscow, 117259, Russia.} 
and Alex Kovner\thanks{E-mail: a.kovner1@physics.ox.ac.uk.}
}

\address{
Theoretical Physics, University of Oxford \\
1 Keble Road, Oxford, OX1 3NP, UK.
}

\date{\today}

\maketitle

\begin{abstract}
We analyze the instanton transitions in the framework of the gauge
invariant variational calculation in the pure Yang-Mills theory.
Instantons are identified with the saddle points in the integration
over the gauge group which projects the Gaussian wave functional onto
the gauge invariant physical Hilbert space. We show that the dynamical
mass present in the best variational state provides an infrared cutoff
for the instanton sizes. The instantons of the size $\rho<1/M$ are
suppressed and the large size instanton problem arising in the standard
WKB calculation is completely avoided in the present variational framework.
\end{abstract}


\section{Introduction}

In recent years there have been renewed interest in application of the
Hamiltonian methods to the study of nonabelian Yang-Mills theories
\cite{freedman,KK95,mansf,nair,iancu}.  One set of these works
\cite{freedman,mansf,nair} attempts to solve ``exactly'' the strongly
interacting gauge theories in the sense that a nonlinear transformation
is performed to a set of gauge invariant coordinates. One then tries to
find a controlled expansion akin to strong coupling perturbation
theory, which hopefully solves the infrared part of the theory in the
leading order.  Another set \cite{KK95,iancu} attempts to find the
Yang-Mills vacuum wave functional with the help of a variational
approximation. In particular in \cite{KK95} a gauge invariant
generalization of a Gaussian variational approximation was developed.
The hope of this approach is that the vacuum of QCD may be not very
different from the vacuum of a free theory in many important respects.
This hope rests on the observation that many genuine nonperturbative
effects in QCD appear already on the momentum scales much larger than
$\Lambda_{\rm QCD}$ where the coupling constant is still small \cite{svz}.
It may be possible then to account for these effects with the Gaussian
wave functional, which is similar to the ground state of the free
theory modifying only the width of the Gaussian for the low momentum
modes.  This modification is essentially nonperturbative and should
lead to the generation of the same condensates that account for a
variety of QCD phenomenology via the QCD sum rules \cite{svz}.

The variational calculations of \cite{KK95} are of exploratory nature
and many questions regarding their validity remain unsettled. Some of
those are discussed in the original paper \cite{KK95} as well 
as in \cite{diakonov}.
Nevertheless 
this variational approach captures many of the
essential features of gluodynamics; mass generation, formation of the
gluon condensate, \cite{KK95}, and asymptotic freedom,
\cite{BK97,B97,diakonov}.  Possible appearance of the linear potential
between static quarks in this approach has also been discussed
\cite{Z98,diakonov,kosv}.  In fact one of the nicer features of this
approximation is that it exhibits nontrivial nonperturbative infrared
physics (gluon condensate) along with correct weak coupling ultraviolet
behavior (one loop Yang-Mills $\beta$ function).  If so one is
naturally lead to ask whether it also gives a good account of the
instanton physics.  Instantons are the only concrete description of the
non-perturbative nature of the QCD vacuum in the path integral
formalism.  In the ultraviolet region although the effect of the
instantons is nonperturbatively small they are easily identifiable.
They should therefore serve as a useful probe for any non-perturbative
approach especially if it purports to capture both the infrared and the
ultraviolet physics.  The aim of the present paper is precisely to
study the structure and the properties of the instantons in the
variational approach of \cite{KK95}.

Instantons are localized, finite-action classical solutions of the
field equations of QCD in Euclidean space-time \cite{bpst}. Such
solutions have been obtained in exact analytical form, and extensively
studied.  An excellent review of instantons in gauge theories can be
found in \cite{S94} and \cite{R87}.

Physically instantons  represent tunneling processes between
topologically distinctive  vacuum sectors with the exponent of the
instanton action being equal to the transition probability between two
of these vacuum states.  As any tunneling probability at weak coupling
it is nonperturbatively small (of order $\exp(-{const}/\alpha_S)$)
and the instantons therefore are invisible in the weak coupling
perturbation theory.  When first discovered there was hope that
instantons would provide the solution to the strong coupling
problem\cite{CDG78,CDG79}.  Although it has been subsequently realized
that instantons are irrelevant for understanding confinement, they have
provided a beautiful mechanism of spontaneous chiral symmetry breaking
\cite{D95}.  The instanton liquid model (initially introduced on
phenomenological grounds \cite{S82}, and later justified by Euclidean
variational methods \cite{DP84}) to this day remains the most complete
theory of chiral symmetry breaking in QCD.  It is therefore vital to
understand the properties of the instantons if one is hoping to extend
the application of the Gaussian variational approximation to QCD with
fermions.

Although the notion of the instanton is intrinsically Euclidean, the
tunneling between different vacuum sectors can be formulated in the
Hamiltonian language as well as in the Lagrangian one.  In fact the
gauge projected formalism of \cite{KK95} is very well suited for this
purpose. The projection of the initial Gaussian onto the gauge
invariant subspace is achieved by the integration over the gauge group.
As will be explained in detail later, the $SU(N)$ matrix $U(x_i)$ (the
dependence here is on the spatial coordinates only) in this type of
calculation turns out to play the role of a $\sigma$-model field.
This field is governed by an ``action'', whose structure depends on the
parameters of the variational state. As we shall see, the Yang-Mills
instantons correspond to the topologically nontrivial saddle points of
this action. In this paper we will study the properties of these saddle
point solutions.

Our main result is the following. We find that the appearance of the
dynamical mass parameter in the variational state stabilizes the size
of the instantons.  Recall, that the main unsolved problem of the
dilute instanton gas approximation is that (when account is taken of
the one loop running of the coupling constant) the path integral is
dominated by the large size instantons. The measure for the integration
over the sizes diverges as a power in the infrared and this divergence
renders the dilute gas calculation meaningless. It is usually assumed
that this infrared divergence is eliminated by some nonperturbative
effects.  This is precisely what we find in the variational approach.
The dynamical mass $M$ present in the best variational state suppresses
instantons of sizes $\rho>1/M$.  The size of the stable instanton that
we find turns out to be consistent with the average size of the
instanton in the instanton liquid model.

This paper is structured as follows. In section \ref{sec:basics} we
briefly recall the formalism and the results of \cite{KK95}. We explain
how to identify the tunneling transition in this framework and what
type of classical configuration should be identified with the
instanton. It is also noted that the generation of the dynamical mass
itself found in \cite{KK95} can be directly interpreted in terms of the
condensation of these instantons.

In section \ref{sec:small} we study numerically the structure and the
action of the small size instantons. For these instantons the presence
of the dynamical mass is irrelevant and this calculation is performed
at zero mass.  The profile function of the instanton and the value of
the transition probability are approximately determined by a
variational method. We find that the tunneling probability indeed
scales as $\exp\{-c\alpha_s\}$ with the value of $c$ approximately
two times larger than in the standard Euclidean calculation.  We
explain why this discrepancy is not unexpected in a variational
calculation.

In section \ref{sec:mass} corrections to the solution and the action
due to a non-zero mass gap are calculated.  It is found that the
instanton is stabilized to a size of the order of the inverse mass
scale. This size is directly comparable with that found in the
instanton liquid model.

Finally, section\ref{sec:discussion} is devoted to discussion of our
results and their relation with the instanton liquid model.

\section{The Hamiltonian portrait of an instanton}
	\label{sec:basics}

We start with a brief description of the gauge invariant Gaussian
approximation of \cite{KK95}.  The Ansatz for the QCD wavefunctional
considered in \cite{KK95} is,
\begin{equation}
\Psi[A_i^a]= \int DU(x_i)\Psi_U[A],
	\label{an}
\end{equation}
with,
\begin{equation}
\Psi_U[A]=	\exp\left\{-{1\over 2}\int d^{3}x d^{3}y 
	\ A_i^{Ua}(x)G_{ij}^{-1ab}(x-y)\ A_j^{Ub}(y)\right\}.
	\label{an1}
\end{equation}
Here $A^{U}$ is the gauge transform of the vector potential with
the gauge transformation $U$.
\begin{equation}
A_i^{Ua}(x) =
	S^{ab}(x) A_i^b(x) + \lambda_i^a(x) ,
\end{equation}
with,
\begin{eqnarray}
S^{ab}(x)& =&
\frac{1}{2}{\rm tr} \left[ \tau^a U^\dagger(x) \tau^b U(x) \right] ,\\
\lambda_{i}^a(x) &= &\frac{i}{g} {\rm tr} \left[ \tau^a U^\dagger(x)
\partial_{i} U(x) \right]\nonumber.
\end{eqnarray}
The SU(N) generators are taken to satisfy
the following algebra and normalization,
\begin{equation}
[\tau^a , \tau^b] = 2 i f^{abc}\tau^c , \;\;\; 
	\frac{1}{2} {\rm tr}[\tau^a \tau^b] = \delta^{a b} .
\end{equation}
The state thus constructed obeys Gauss' law since it is explicitly
invariant under the gauge transformation $A_i(x)\rightarrow A_i^V(x)$
with arbitrary $SU(N)$ matrix $V(x)$.  The width of the Gaussian $G(x)$
is a parameter with respect to which the expectation value of the
Hamiltonian is varied.  A simple rotation and (global) color invariant
form is $G_{ij}^{-1ab} = \delta_{ij} \delta^{ab} G^{-1}$.  The
functional form of $G^{-1}$ is chosen to agree with the perturbation
theory in the limit of high momentum on one hand, and to allow for the
nonperturbative mass scale on the other,
\begin{eqnarray}
G^{-1}(k) = \left\{ \begin{array}{ll} \sqrt{ k^{2} ~} & 
	\mbox{ if  $ k^2>M^2$}\\
 	M &  \mbox{ if $k^2<M^2$} 
	\end{array} 
	\right. .
	\label{an4}
\end{eqnarray}

Technically, the calculation of the expectation values of gluonic
operators in the state eq.(\ref{an}) is mapped into the calculation in
a nonlocal nonlinear $\sigma$-model in three Euclidean dimensions.
Consider the vacuum average of an arbitrary gauge invariant operator
$O[A]$,
\begin{eqnarray}
\langle O \rangle &=& \frac{1}{Z}\int DA \Psi^*[A]O[A]\Psi[A]=
{1\over Z}\int DU_1DU_2DA \Psi_{U_1}^*[A]O[A]\Psi_{U_2}[A]\nonumber\\
&=&{1\over Z}\int DUDA \Psi_{1}^*[A]O[A]\Psi_{U}[A] .
\end{eqnarray}
In the last line the matrix $U=U^\dagger_1U_2$ is the relative gauge
transformation between the two Gaussian wave functions.  The
normalization factor $Z$ (the norm of
the state) is, 
\begin{equation}
Z=\int DUDA \Psi_{1}^*[A]\Psi_{U}[A] .
\end{equation}
The Gaussian integration over the gauge potential
$A_i$ can be performed explicitly. As a result the last step of the
calculation is a path integral over the $SU(N)$ matrix $U$ with the
$\sigma$-model partition function, 
\begin{equation}
Z=\int DU\exp\{-\Gamma[U]\} , 
\end{equation}
where the (nonlocal) action is, 
\begin{equation}
\Gamma [U] = \frac{1}{2} \lambda \Delta \lambda +
\frac{1}{2} {\rm Tr} {\rm ln} {\cal M} .	
	\label{q1}
\end{equation}
Here summation over all indices (rotational, color and coordinate)
is implied. The first term is
written explicitly as,
\begin{equation}
\frac{1}{2}\int d^3x \ d^3y \ \lambda_i^a(x) 
	\Delta^{a c}(x,y) \lambda_i^c(y) ,
	\label{axshun}
\end{equation}
with,
\begin{equation}
\Delta^{ac}(x,y) = 
	[\delta^{ac} G(x-y) +S^{ab}(x) G(x-y) S^{cb}(y)]^{-1}.
	\label{invop}
\end{equation} 
In eq.(\ref{q1}) we have defined,
\[
{\cal M}^{ab}_{ij}(x,y) = 
	\left[ S^{Tac(x)} S^{cb}(y) + \delta^{ab} \right] \,
	G^{-1}(x-y) \, \delta_{ij} . 
\]
The second term in eq.(\ref{q1}) is of $O(g^2)$ relative to the first
one and with the accuracy of \cite{KK95} can be ignored. We will not
consider it in the following.  The action of the $\sigma$-model
eq.(\ref{axshun}) depends on the variational parameter $M$ through
eq.(\ref{invop}).  The minimization of the expectation value of the
energy in \cite{KK95} leads to a nonzero value of the mass parameter
$M$ in the best variational state. The dynamical mass parameter is
determined by the relation
\begin{equation}
\alpha_s(M_0)={\pi\over 4N_c} ,
\end{equation}
where the Yang-Mills coupling constant $\alpha_s$ evolves according to
one loop $\beta$-function\footnote{More accurately, the
$\beta$-function in the variational calculation of \cite{KK95} is
slightly different from the complete one loop expression. See
discussion in section \ref{sec:mass}.}.  The significance of this value
of $M$ from the point of view of the effective $\sigma$ model is that
at this point it undergoes the phase transition.  For $M<M_0$ the model
is in the weakly coupled ordered phase. In this phase the matrix $U$ is
close to the unit matrix with small fluctuations around it. For $M>M_0$
the model is in the disordered phase. The matrix $U$ fluctuates
strongly so that it covers all available phase space and its average
value vanishes. It was found in \cite{KK95} that the energy is
minimized just above the phase transition $M=M_0+$ so that the
$\sigma$-model is in the disordered phase.

Coming back to the subject of the present paper, the first thing is to
understand how do we expect to see instantons in this formalism.  The
answer to this is the following. As explained above the effective
$\sigma$-model arises as an integration over the relative gauge
transformations between the two Gaussian states in the linear
superposition eq.(\ref{an}). The Boltzmann factor $\exp(-\Gamma[U])$ for
a given matrix $U$ is therefore just the overlap of the initial and the
gauge rotated state, or in other words the transition amplitude between
the two states\footnote{Since the space of the matrices $U$ is
continuous, strictly speaking the Boltzmann factor is the differential
rather than the total amplitude.}.  The instanton transition is
precisely the transition of this type, where the two states are related
by a large gauge transformation.  The matrix of this large gauge
transformation must carry a nonzero topological charge $\Pi_3
(SU(N))$.

The integration measure over $U$ indeed includes integration over
topologically nontrivial configurations.  The finiteness of the action
eq.({\ref{axshun}) requires that the matrix $U$ approaches constant
value at infinity.  This identifies all points at spatial infinity
hence, the physical space of the model is $S^3$.  Field configurations
are maps from $S^3$ into the manifold of $SU(N)$ and are classified by
their winding number, or topological charge, which is an element of the
homotopy group $\Pi_3 (SU(N)) = Z$.  The $\sigma$-model action in a
given topological sector is minimized on some configuration which is a
solution of classical $\sigma$-model equations of motion. In
particular, the solution with a unit topological charge is expected to
have a ``hedgehog'' structure much like the topological soliton in the
Skyrme model\cite{B94}.  The integral over $U$ in the steepest descent
approximation is saturated by these classical solutions.

These $\sigma$-model configurations that belong to a nontrivial
topological sector with a unit winding number represent QCD transitions
between the topologically distinct sectors.  The topologically
nontrivial classical soliton solutions of the $\sigma$-model are
therefore the three dimensional images of the QCD instantons.

The QCD instantons are defined in space time and are therefore four
dimensional point-like objects. The $\sigma$-model solutions are
intrinsically three dimensional.  Nevertheless, there is a natural
simple relation between the two. For a given Yang-Mills instanton
solution $A^{\rm inst}(x_\mu)$ one can find a three dimensional $SU(N)$
matrix $U(x_i)$ by the procedure discussed by Atiyah and Manton,
\cite{AM89},
\begin{equation}
U_{\rm AM}(x_i)=P\exp \left( i\int_Cdx^\mu A^{\rm inst}_\mu \right) ,
\end{equation}
where the contour of integration $C$ is a straight line $x_i=const$,
$-\infty<x_0<\infty$.  The matrix $U_{\rm AM}$ gives the relative gauge
transformation between the initial trivial vacuum at
$x_0\rightarrow-\infty$ and the topologically nontrivial vacuum at
$x_0\rightarrow +\infty$, or in other words between the initial and
final states of the instanton transition.  Clearly, its meaning is
precisely the same as of the classical soliton solution of the
effective $\sigma$-model eq.(\ref{axshun}).  Also, the QCD instanton
action and the $\sigma$-model soliton action have the same physical
meaning. They both give the transition probability between different
topological sectors in QCD.  We will therefore refer to the
$\sigma$-model solitons as instantons in the following.

One has to realize that although the QCD  and the $\sigma$ - model
instantons have the same physical meaning, 
it is not assured that the numerical value for
their respective actions is the same.
They both approximate the value of the transition probability in QCD,
but the approximations involved are quite different. The QCD instanton
action is the result of the standard WKB approximation which is
valid at weak coupling and therefore for small instantons, but breaks
down for instantons of large size. The $\sigma$ - model instanton
action on
the other hand is the value of this transition probability in a
particular Gaussian variational approximation. It is natural to expect
that variational calculation underestimates the value of the
transition probability at very weak coupling.
The transition probability is given by the
overlap of the ``ground state'' wave functions in two topological
sectors. For simplicity let us consider a quantum mechanical system
with two vacua at $x_\pm$. If the area below the barrier separating the
vacua is large, the standard WKB instanton calculation is applicable. The
wave function of each of the vacua below the barrier has essentially an
exponential fall off $\exp\{i\int^x \sqrt{E-V(x-x_\pm)}\}$. The
instanton calculation is the calculation of the overlap of these
functions.  Our variational calculation corresponds to approximating
the respective ``ground states'' at $x_\pm$ by Gaussian wave functions.
The tails of the Gaussians fall off much faster away from the minimum
than the actual wave function and the overlap is therefore is expected
to be smaller. 
When the coupling constant is not too small
(or when the area below the barrier is not too large) the overlap between the
two states is not determined any more by the behavior of the ``tails'' of
the wave functions. In this situation one can expect the Gaussian
approximation to do much better, since the overlap region contributes
significantly to the energy and therefore plays important role in the
minimization procedure.

The rest of this paper is devoted to a quantitative study of instantons
in the variational state eq.({\ref{an}).  Before proceeding to this
part, however, we would like to note that implicitly the instantons
played a very important role already in the energy minimization of
\cite{KK95}. As we mentioned above the energy is minimized for the
value of the mass parameter $M$ at which the $\sigma$ model is in the
disordered phase. The transition between ordered and disordered phases
in a statistical mechanical system can usually be described as a
condensation of topological defects. This is a standard description of
the phase transition in the Ising and XY models \cite{ising}.  In the
$\sigma$ model eq.(\ref{axshun}) the relevant topological defects are
none other than the instantons. In this sense the appearance of the
dynamical mass in the best variational state itself is driven by the
condensation of instantons.

\section{Small Size Instantons} \label{sec:small}

In this section we wish to study small size instantons.  For the
instanton solutions of a size $\rho<<1/M$ the presence of a finite
dynamical mass is irrelevant.  We will therefore take $M=0$ for the
calculations in this section.  The existence of a finite mass scale is
very important for the instantons of large size and will be taken into
account in the next section.  It somewhat increases the complexity of
the calculation but all the necessary methods can be developed for the
$M=0$ case.

For $M=0$ the inverse propagator (\ref{an4}) which defines the
variational state eqs.(\ref{an},\ref{an1}) is (in momentum space)
$G^{-1}(k)=|k|$. We find it more convenient to work in coordinate space
throughout the rest of this paper.  Fourier transforming $G^{-1}$ to
the coordinate space according to the definition,
\begin{equation}
G^{-1}(k)= (2 \pi)^{-3/2} \int d^3x G^{-1}(x) e^{ikx},
\end{equation}
we find,
\begin{equation}
G^{-1}(x-y) = -\frac{1}{\pi^2} 
	\left( \frac{\Theta(|x-y|-\Lambda^{-1})}{|x-y|^4}
	- \Lambda^3 \delta(|x-y|-\Lambda^{-1}) \right).
	\label{g1old}
\end{equation}
Here we had to introduce the ultraviolet cutoff $\Lambda$ to define the
coordinate space expression properly.  The coefficient of the second
term is determined by requiring that at finite cutoff $\Lambda$,
\begin{equation}
\int dx^3G^{-1}(x)\propto G^{-1}(k=0)=0 .
\end{equation}
Note that $\Lambda$ is introduced as a regulator only and
no subtraction in the expression for the propagator (or the action)
was performed. 
The cutoff $\Lambda$ should be taken to infinity at the end of the
calculation and the results of the calculation should be finite in
this limit. Below we show how the divergent terms cancel exactly, so
that in the numerical calculations we only take into account finite
terms in the limit $\Lambda \to \infty$.

We are searching for an instanton solution of topological charge one to
the action (\ref{axshun}).  Such a solution should have the maximally
symmetric ``hedgehog'' form,
\begin{equation}
U(x) = e^{i \tau^a \hat{x}^a f(r)} = 
	\cos{f(r)} + i \tau^a \hat{x}^a \sin{f(r)} ,
	\label{skyrmion}
\end{equation}
where $\tau^a$ are the generators of $SU(2)$ and $f(r)$ is an
unspecified function of $r = |x|$, which we will refer to 
as the profile function.
The profile function 
is constrained to satisfy $f(0) = \pi$ and $f(\infty) = 0$,
which ensures that the field configuration 
eq.({\ref{skyrmion}) has unit topological charge.  
We need only consider the group $SU(2)$ since the solutions for
$SU(N>2)$ can be found from the embedding of $SU(2)$ in $SU(N)$,
\cite{B79}. Substituting this form for the field into the components of
the action, we find,
\begin{eqnarray}
\lambda_i^a(x) &=& 
	\frac{i}{g} {\rm tr}[\tau^a U^+(x) \partial_i U(x)]
	\\ \nonumber
	&=& -\frac{2}{g} \hat{x}^i \hat{x}^a f'(r) - 
	\frac{1}{gr} (\delta^{ia} - \hat{x}^i \hat{x}^a) \sin{2 f(r)}
	-\frac{2}{g} \epsilon_{abi}\frac{\hat{x}^b}{r} \sin^2 f(r), 
	\\
S^{ab}(x) &=& \frac{1}{2} {\rm tr}[\tau^a U^+(x) \tau^b U(x)] 
	\\ \nonumber
	&=& \delta^{ab} - \epsilon_{abc}\hat{x}^c \sin{2 f(r)} + 
	2 (\hat{x}^a \hat{x}^b - \delta^{ab}) \sin^2 f(r) ,
\end{eqnarray}
where $\epsilon_{abc}$ is the antisymmetric tensor.  For the hedgehog
configuration, 
the function $S^{ab}(x)$ differs significantly 
from $\delta^{ab}$ only over a small
region, the size of which depends on how fast the transition is from
the asymptotic behavior at large distance to that at small distance.
If the profile function has a sharp transition between these two limits
one can reasonably  approximate $S^{ab}$ by $\delta^{ab}$.  Once the
form of the profile function has been found this can be checked for
consistency.  It is possible to expand around $S^{ab}=\delta^{ab}$ and
systematically calculate corrections in $S^{ab}-\delta^{ab}$.
 At the end of this section we calculate the
leading correction and find that it is indeed small for the instanton profile
function. 

This approximation allows us to write,
\begin{equation}
\Delta^{ac}(x,y) =
[\delta^{ac} G(x-y) + S^{ab}(x) G(x-y) S^{cb}(y)]^{-1} \simeq
\frac{1}{2} G^{-1}(x-y) \delta^{ac} .
\label{acap}
\end{equation}

With the definition eq.(\ref{g1old}), we write the action
eq.(\ref{axshun}) as,
\begin{eqnarray}
\Gamma &=& \Gamma_1 - \Gamma_2 , \\ \nonumber
\Gamma_1 &=& - \frac{1}{4\pi^2} \int d^3x d^3y \lambda_i^a(x)
\frac{\Theta(|x - y| - \Lambda^{-1})}{|x - y|^4} \lambda_i^a(y) ,
\\ \nonumber
\Gamma_2 &=& -\frac{\Lambda^3}{4\pi^2} \int d^3x d^3y \lambda_i^a(x)
\delta(|x-y| - \Lambda^{-1}) \lambda_i^a(y) .
\end{eqnarray}
Both $\Gamma_1$ and $\Gamma_2$ are divergent. The divergences between
the two terms however cancel leaving the action finite. Changing
variables, $u_i=x_i+y_i$, $v_i=x_i-y_i$, and using the fact that $2
|v|^{-4} = \nabla^2_v |v|^{-2}$, we find after integrating by parts
twice,
\begin{eqnarray}
\Gamma_1 &=& -\frac{1}{64\pi^2} \int d^3u d^3v
	|v|^{-2} \nabla^2_v \left[
	\lambda_i^a(x)
	\lambda_i^a(y) \right] + \Gamma_{\rm surf} 
	+ \Gamma_{\rm div} , \\
\Gamma_{\rm surf} &=& \left.
	- \frac{1}{64\pi^2} \int d^3u d\Omega_v 
	\left( 2 |v|^{-1} + \partial_{|v|} \right)
	\left[
	\lambda_i^a(x)
	\lambda_i^a(y) \right]
	\right|_{|v| = \infty} , \\
\Gamma_{\rm div} &=& \left.
	\frac{1}{64\pi^2} \int d^3u d\Omega_v 
	\left( 2 |v|^{-1} + \partial_{|v|} \right)
	\left[
	\lambda_i^a(x)
	\lambda_i^a(y) \right]
	\right|_{|v| = \Lambda^{-1}} ,
	\label{actq}
\end{eqnarray}
where $d\Omega_v$ denotes angular integration in $v$-space, and
$\nabla^2_v$ is the $v$-space Laplacian.  The divergent part
$\Gamma_{\rm div}$ cancels exactly the ``subtraction'' term $\Gamma_2$
when $\Lambda \rightarrow \infty$.  Moreover, the surface terms
$\Gamma_{\rm surf}$ vanish if the profile $f(r)$ approaches zero as
$r^{-2}$ or faster at infinity, and is infinite otherwise.  In what
follows we assume that $f(r)$ decreases fast enough. 
Ignoring all terms in the action which vanish in the limit 
$\Lambda \to \infty$ we are left with the finite, cutoff independent action,
\begin{equation}
\Gamma = - \frac{1}{8 \pi^2} \int d^3x d^3y
	\frac{1}{|x-y|^2} \left(\frac{\partial^2}{\partial x_j^2} 
	+ \frac{\partial^2}{\partial y_j^2} - 
	2 \frac{\partial^2}{\partial x_j \partial y_j} \right) 
	\left[\lambda_i^a(x) \lambda_i^a(y) \right] .
	\label{acfin}
\end{equation}
For the hedgehog configuration (\ref{skyrmion}),
\begin{eqnarray}
\lambda_i^a(x) \lambda_i^a(y) &=& 
	\frac{1}{g^2} \left[ 4 \cos^2 \theta f'(r) f'(s) + 
	\frac{2}{s} (1 - \cos^2 \theta) f'(r) \sin{2 f(s)} \right.
	\label{lili}
	\\
	& & +
	\frac{2}{r} (1 - \cos^2 \theta) f'(s) \sin{2 f(r)} +
	\frac{1}{rs} (1+ \cos^2 \theta) \sin{2 f(r)} \sin{2 f(s)} 
	\non \\
	& & + \left. \frac{8}{rs}\cos \theta \sin^2 f(r) \sin^2 f(s) 
	\right],	
	\non
\end{eqnarray}
where $\cos \theta = \hat{x} \cdot \hat{y}$, $r=|x|$ and $s = |y|$. From
(\ref{lili}) we see that $\lambda_i^a(x) \lambda_i^a(y)$ is a function
of only three variables, $\lambda_i^a(x) \lambda_i^a(y) = \frac{1}{g^2}
\sum_{n=0}^2 \cos^n \theta \, H_n(r,s)$, where $H_n$ are immediately found
from (\ref{lili}). They are given explicitly in the Appendix for easy
reference.  We also give in the Appendix the coefficient functions
$\tilde H_n(r,s)$ which are defined by,
\begin{equation}
\left(\frac{\partial^2}{\partial x_j^2} 
	+ \frac{\partial^2}{\partial y_j^2} - 
	2 \frac{\partial^2}{\partial x_j \partial y_j} \right) 
	\left[\lambda_i^a(x) \lambda_i^a(y) \right] = \frac{1}{g^2} 
	\sum_{n=0}^3 \cos^n \theta \, \tilde{H}_n(r,s) .
\end{equation}

After carrying out the angular integrations we get an action
as a functional of the profile function, 
\begin{equation}
\Gamma = - \frac{2}{g^2} \int dr \, ds \, (r s)^2 \sum_{n=0}^3 I_n(r,s) 
	\tilde{H}_n(r,s),
	\label{action0}
\end{equation}
where,
\begin{equation}
I_n(r,s) = \int_{-1}^1 \frac{d(\cos{\theta}) \cos^n \theta}{(r^2 + s^2
	- 2 r s \cos \theta)} ,
	\label{int}
\end{equation}
are given in the Appendix.

We were unable to minimize the action with respect to the profile
function by analytical methods. We have therefore employed a variational
method to determine the best profile function approximately.  The two
important variational parameters are the ones that govern the
asymptotic behavior of $f(r)$,
\begin{equation}
\left\{
	\begin{array}{lcl}
	f \sim r^{-\alpha} & & r \rightarrow \infty \\
	f \sim \pi - r^{\beta} & & r \rightarrow 0 \\
	\end{array}
	\right. .
\end{equation}

The dependence on the parameter $\alpha$ which determines the
asymptotics at large distance turns out to be simple. Using two trial
functions,
\begin{eqnarray}
f_1(r) &=& \pi \left[ 
	\frac{\rho^{\alpha}}{\rho^{\alpha} + x^{\alpha}} \right] , \\
f_2(r) &=& 2 \arctan{(\rho/r)^{\alpha}} , 
\end{eqnarray}
and performing the double numerical integration we have found that the
action monotonically increases with $\alpha$.  This means that the
optimal value for this parameter is $\alpha = 2$, since this is the
lowest possible value at which the surface terms are non-infinite.

Having fixed $\alpha$, we next studied the dependence on the
parameter $\beta$ which determines the behavior close to the
instanton center.  The trial functions we used for this purpose are,
\begin{eqnarray}
f_1(r) &=& \pi \left[ 
	\frac{\rho^{\beta}}{\rho^{\beta} + 
	x^{\beta}} \right]^{2/\beta} , 
	\label{f1} \\
f_2(r) &=& 2 \arctan \left( \frac{2}{(r/\rho)^2 + 
	(r/\rho)^{\beta}} \right) . 
	\label{f2}
\end{eqnarray}
We find numerically the optimal value for the parameter $\beta$ to be
$\beta = 1.1$ for $f_1$ and $\beta = 1.0$ for $f_2$.

The corresponding values for the action are
\begin{equation}
\Gamma = 1.75 \frac{8 \pi^2}{g^2},
\end{equation} 
for $f_1$, and 
\begin{equation}
\Gamma = 1.73 \frac{8 \pi^2}{g^2},
\end{equation} 
for $f_2$. 

We have calculated the first 
correction $\Delta\Gamma$ to the action in the expansion of $S^{ab}$ around
$\delta_{ab}$ for the two profiles. We find
$\Delta\Gamma=0.21 \frac{8 \pi^2}{g^2}$ for $f_1$ and 
$\Delta \Gamma=0.34 \frac{8 \pi^2}{g^2}$ for
$f_2$. 
The correction to the action for both profiles is of order of
$10\%$. Note however, that the Ansatz $f_1$ is more stable, and indeed
after the correction is taken into account has lower action than $f_2$.

The corrected (to this order) values
for the action are
\begin{equation}
\Gamma = 1.96 \frac{8 \pi^2}{g^2},
\label{prob}
\end{equation} 
for $f_1$, and 
\begin{equation}
\Gamma = 2.07 \frac{8 \pi^2}{g^2},
\end{equation} 
for $f_2$. 

Note that although our Ansatze for the profile function depend on the
instanton size $\rho$, the value of the action does not depend on it. This is 
the direct consequence of the dilatational invariance of the
effective $\sigma$-model action eq.(\ref{acap}). 
The introduction of the cutoff $\Lambda$ strictly speaking 
breaks the dilatational
invariance. This breaking however is very small and 
the invariance is restored in the limit
$\Lambda\rightarrow\infty$.

We now want to comment on the numerical value of the transition
probability obtained in our variational approach. 
Our result eq.(\ref{prob}) 
should be compared with the value $8 \pi^2/g^2$ for the
classical action of a Yang-Mills instanton.
So the action of an 
instanton in the variational approach 
is about two times larger than the standard path integral
result and the transition probability therefore appears to be much
smaller.  
As explained in Section II
it is in fact natural to expect this sort of behavior in the
Gaussian approximation 
since the tails of the Gaussians fall off much faster away from the minimum
than the actual wave function and the overlap is therefore smaller.
This is the basic reason for the discrepancy in the numerical values of
the action between the variational instanton of this section and the
weak coupling (WKB) instanton of the standard path integral approach.

It is significant that both the asymptotic behavior and the value of
the action for both our variational Ansatze for the profile function 
is very similar to the corresponding results for the Atiyah-Manton 
Ansatz\cite{AM89}.
As discussed in Section II, the natural identification between
the QCD instanton and the instanton in the effective $\sigma$-model 
is furnished by the 
holonomy along all time-lines,
\begin{equation}
U(x) = {\rm T} \exp 
	\left[-\int_{-\infty}^{\infty}dx^\mu A_\mu^{\rm inst}(x)  \right].
\label{inst}
\end{equation}

For the QCD instanton eq.(\ref{inst}) results in the profile
function,
\begin{equation}
f_{\rm AM}(r) = \pi \left[ 1 - 
	\left(1+\frac{\rho^2}{r^2}\right)^{-\frac{1}{2}} 
	\right].
	\label{prof}
\end{equation}
The asymptotics of this configuration at large distances is
$f(r)\propto 1/r^2$, and near the instanton core $f(r)-\pi\propto r$.
This gives $\alpha=2$, $\beta=1$ for the Atiyah-Manton configuration,
identical with $f_2$ and practically indistinguishable from $f_1$. 
In fact the profile $f_{\rm AM}$ is very similar to $f_1$ not only
asymptotically but also in the whole range $0<r<\infty$.
In
Fig. \ref{fig0} we plot the three profile functions for the same value
of the instanton size. 
It is clear that
$f_1$ and $f_{\rm AM}$ are very similar while $f_2$ differs from them
somewhat inside the instanton core. We also note that $f_2$ has a
slower approach to its value at $r=0$, which explains why the value of
the action for this profile is less stable against the
$S^{ab}-\delta^{ab}$ corrections.
 
\begin{figure}
\begin{center}
\leavevmode
\epsfxsize=4.3in
\epsfbox{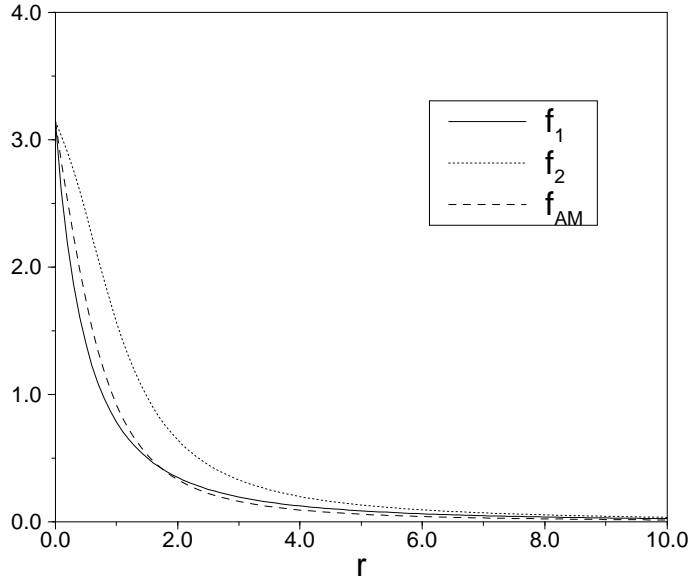}
\caption{The hedgehog profile functions $f_1$, $f_2$ and $f_{\rm AM}$ 
for $\rho = 1$.} 
\label{fig0}
\end{center}
\end{figure}

We have calculated the
$\sigma$-model action for the Atiyah-Manton 
profile numerically and found (including the first correction in
$S^{ab}-\delta^{ab}$ expansion) 
$\Gamma =
1.97 \frac{8 \pi^2}{g^2}$.  This again is very close to the value we
obtain with our variational Ansatz eq.(\ref{prob}).
This means that even though the
value of the transition probability is underestimated in the Gaussian
approximation, the actual field configurations into which the
tunneling is most probable are identified correctly--they are
precisely the same as in the WKB calculation.

It should be noted that the exact value of the
transition probability due to small size instantons (very weak
coupling) is not so important, since they do not significantly affect
the structure of the vacuum in any case. It would therefore be
incorrect to declare the Gaussian approximation as a failure from the
instanton point of view. 
One expects the Gaussian
approximation to do much better at intermediate couplings, 
since the overlap region contributes
significantly to the energy and therefore plays important role in the
minimization procedure. 
In the Yang-Mills framework this means that
there are two types of instantons configurations that
we expect to
contribute significantly and therefore to be better described by the
Gaussian approximation. First, these are large instantons for which
the coupling constant is not small and the action not large. Those are
presicely
the configurations that cause problems in dilute instanton gas approximation.
The second class is the multi-instanton configurations, where the
separation between instantons is not too large and the interaction
between them is not negligible. These are the configurations of the
instanton liquid type. As discussed in the previous section,
these configurations are in fact responsible for the generation
of the dynamical scale in the best variational state through the
condensation of instantons.
The direct discussion of these multi-instanton configuration is however
beyond the scope in the present paper. The large size instantons, or
rather the instantons of the size comparable to the dynamical scale
$1/M$ are the subject of the next section.

\section{Mass Correction and Instantons of Stable Size} 
	\label{sec:mass}

We now want to explore the effect of the dynamical mass on the
properties of the instantons. Our expectation is that the presence of
the mass scale stabilizes the size of the instantons and suppresses
the instantons of the sizes larger than $1/M$. 

The simple qualitative argument to this effect is the
following. Consider the effective $\sigma$-model action for very large
size instantons. In such a configuration only field modes with small
momentum $k<M$ are present. For these momenta the action simplifies,
and as discussed in \cite{KK95} the action eq.(\ref{axshun})
becomes the standard local
$\sigma$-model where $M$ plays the role of the ultraviolet cutoff.
\begin{equation}
\Gamma={1\over 2}{M \over g^2(M)} tr
 \int d^{3}x ~\partial_iU^\dagger(x)\partial_iU(x)
\label{low}
\end{equation}
If the large size instantons are stable at all they should also
be present 
as stable solutions in this local action eq.(\ref{low}).
However this is not the case as can be easily seen by the standard
Derrick type scaling argument.
Take an arbitrary  configuration $u(x)$ in the instanton sector 
and scale all the coordinates
by a common factor $\lambda$. Then obviously
\begin{equation}
\Gamma[u(\lambda x)]=\lambda^{-1} \Gamma[u(x)]
\end{equation}
The dependence of the action on $\lambda$ is monotonic and
is minimized at $\lambda\rightarrow\infty$.
This means that the instantons in the local $\sigma$-model shrink 
to the ultraviolet cutoff $1/M$. For instantons smaller than the
inverse
cutoff we can not use the
local action anymore. However the behavior of these small size 
instantons is already familiar. We know that when the running of the
coupling is taken into account, these instantons are pushed to
the large size. This is the infrared problem of large instantons we
alluded to earlier. 
In our variational state the coupling constant stops running at the
scale $M$. The picture is therefore very simple. The small size
instantons are pushed to larger size by the effect of the coupling
constant, while the large size instantons are pushed to smaller size
by the effect of the local $\sigma$-model scaling. It is therefore
clear that the size will be stabilized somewhere in the vicinity of 
$\rho\sim1/M$.

To confirm this picture we now turn to the numerical evaluation of the
instanton action.
We
write the inverse propagator (\ref{an4}) as,
\begin{equation}
G^{-1}(k) = G_0^{-1}(k) + \Delta G^{-1}(k),
	\label{prop}
\end{equation}
where 
\begin{eqnarray}
G_0^{-1}(k) &=& |k|\\
\Delta G^{-1}(k) &=& \theta(M - |k|) \ (M- |k|)\nonumber
\end{eqnarray}

 Since the action eq.(\ref{axshun}) is linear in $G^{-1}$,
the  $G_0^{-1}$ term will lead to the classical action $\Gamma_0$
calculated in the previous section.
As
discussed above, $\Gamma_0$ is scale invariant and therefore does not
depend on the size of the instanton.
In coordinate space, the second term
in eq.(\ref{prop}) reads,
\begin{equation}
\Delta G^{-1}(|x-y|) = - \frac{1}{\pi^2 |x-y|^4} 
	\left( \cos{M|x-y|} - 1 + \frac{M|x-y|}{2}
	\sin{M|x-y|} \right).
\end{equation}

The mass correction term in the propagator gives rise to the correction
in the action 
\begin{equation}
\Delta \Gamma = \frac{1}{4} \int d^3x d^3y 
	\lambda_i^a(x) \Delta G^{-1}(x-y)\lambda_i^a(y).
	\label{kkmm}
\end{equation}

We now compute the dependence of $\Delta \Gamma$ on the size of the
instanton for the three profile functions considered above, $f_1$,
$f_2$ and $f_{\rm AM}$. As before, the angular integrations can be
performed exactly. We obtain,
\begin{equation}
\Delta \Gamma = - \frac{2}{g^2} 
	\int dr ds (r s)^2 \sum_{n=0}^3 \tilde{I}_n(r,s) 
	H_n(r,s),
	\label{actionm}
\end{equation}
where $H_n$ are as before and $\tilde{I}_n$ defined by
\begin{equation}
\tilde{I}_n = \int_{-1}^1 
	\frac{d(\cos{\theta}) \cos^n \theta}
	{(r^2 + s^2 - 2 r s \cos \theta)^2} 
	\left( \cos{M|x-y|} - 1 + \frac{M|x-y|}{2}
	\sin{M|x-y|} \right) ,
	\label{intm}
\end{equation}
are given in the Appendix.

We calculated numerically the dependence of the classical
action\footnote{In this section we limit ourselves to the approximation
$S^{ab}=\delta^{ab}$. The corrections to this approximation 
as we saw in the previous
section are fairly small and we will not explore them here.} on the
size of the instanton for the three profiles (\ref{f1}), (\ref{f2}) and
({\ref{prof}). We find that the action increases monotonically with
the instanton size. The dependence of the classical action on the size
for the three anzatse is shown in Fig. \ref{fig1}.
As discussed in the beginning of this section the classical action
favors small size instantons in all three Ansatze.

\begin{figure}
\begin{center}
\leavevmode
\epsfxsize=4.3in
\epsfbox{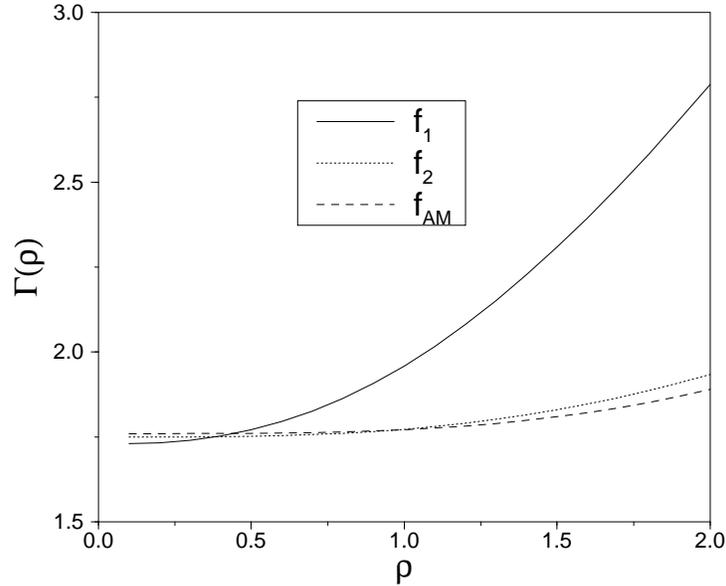}
\caption{Dependence of the classical action on the size of the
instanton for the profiles $f_1$, $f_2$ and $f_{\rm AM}$. The action is
given in units of $8 \pi^2/g^2$.} \label{fig1}
\end{center}
\end{figure}

We now must include the effect of the running coupling constant.
This is equivalent to calculating the one loop correction around the
instanton background.
Rather than performing this technically nontrivial calculation we will
use the results of \cite{BK97}. We will limit ourselves to the $SU(2)$
theory in the following. The $\beta$-function in the nonlinear
$\sigma$-model was found in \cite{BK97} as
\begin{equation}
\beta(g) = -\frac{g^3}{(4 \pi)^2} 8 .
\label{betaf}
\end{equation}
This is slightly different from the complete one loop Yang-Mills
$\beta$-function. The origin of this 
discrepancy was studied in \cite{BK97,B97} and is
well understood now. It is due to the fact that the present
variational calculation omits the screening contributions of the
transverse gluons.
As discussed in \cite{diakonov} 
the variational state can be modified to yield an
exact
$\beta$-function. This point is not essential for our analysis and we
will not dwell on it any 
further except noting that the use of either eq.(\ref{betaf}) or the
exact one loop expression in the following leads to the same results.

Eq.(\ref{betaf}) is valid at distances smaller than $1/M$. The
coupling constant at these distances therefore scales according to
\begin{equation}
\frac{8 \pi^2}{g^2(\mu>M)} = 8 \ln \left(
	\frac{\mu}{\Lambda_{\rm QCD}} \right) ,
	\label{runc}
\end{equation}
At distances larger than $M$ the running of the coupling constant stops
and it tends to a constant value,
\begin{equation}
\frac{8 \pi^2}{g^2(\mu<M)} = 8 \ln \left(
	\frac{aM}{\Lambda_{\rm QCD}} \right) .
	\label{runc2}
\end{equation} 
with $a$, a numerical constant of order one.
The precise interpolation between the two regime is unimportant and we
will use the following simple interpolating expression
\begin{equation}
\frac{8 \pi^2}{g^2(\mu)} = 4 \ln \left( 
	\frac{M^2}{\Lambda_{\rm QCD}^2} 
	\left(\frac{\mu^2}{ M^2} + a \right) \right) .
	\label{runca}
\end{equation}
Since the exact value of the constant $a$ is not known we will present
our results for several values of order one.
When evaluating the action of the instanton of the size $\rho$ we
obviously must take $\mu=1/\rho$.

The running of the coupling constant is not the only logarithmic
effect at one loop order. In addition one has to take into account
the path integral measure over the three translational zero
modes and the size of the instanton,
\begin{equation}
dx_1 dx_2 dx_3 d\rho \rho^{-4} , 
\end{equation}
which contributes an extra term to the action,
\begin{equation}
\Gamma_{\rm measure} = 4 \log{\rho} .
\end{equation}
Collecting all the one-loop logarithmic contributions together we
calculate the corrected action as a function of the size of the
instanton as
\begin{equation}
\Gamma(g^2(1/\rho))+\Gamma_{\rm measure} .
\end{equation}

The result for the three different profiles we have
considered are shown in Figs. \ref{fig2}, \ref{fig3} and \ref{fig4},
for different values of the parameter $a$. 

\begin{figure}
\begin{center}
\leavevmode
\epsfxsize=4.3in
\epsfbox{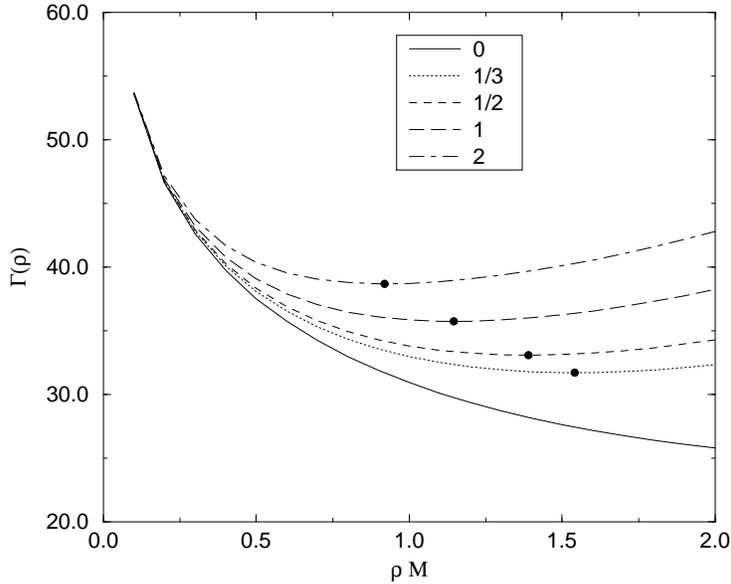}
\caption{The classical action plus one-loop logarithmic corrections for
the profile $f_1$ as a function of the instanton size. The different
curves correspond to different values of the parameter $a = 0, \ldots,
2$, and the bullets mark the minimum of each curve.} \label{fig2}
\end{center}
\end{figure}

\begin{figure}
\begin{center}
\leavevmode
\epsfxsize=4.3in
\epsfbox{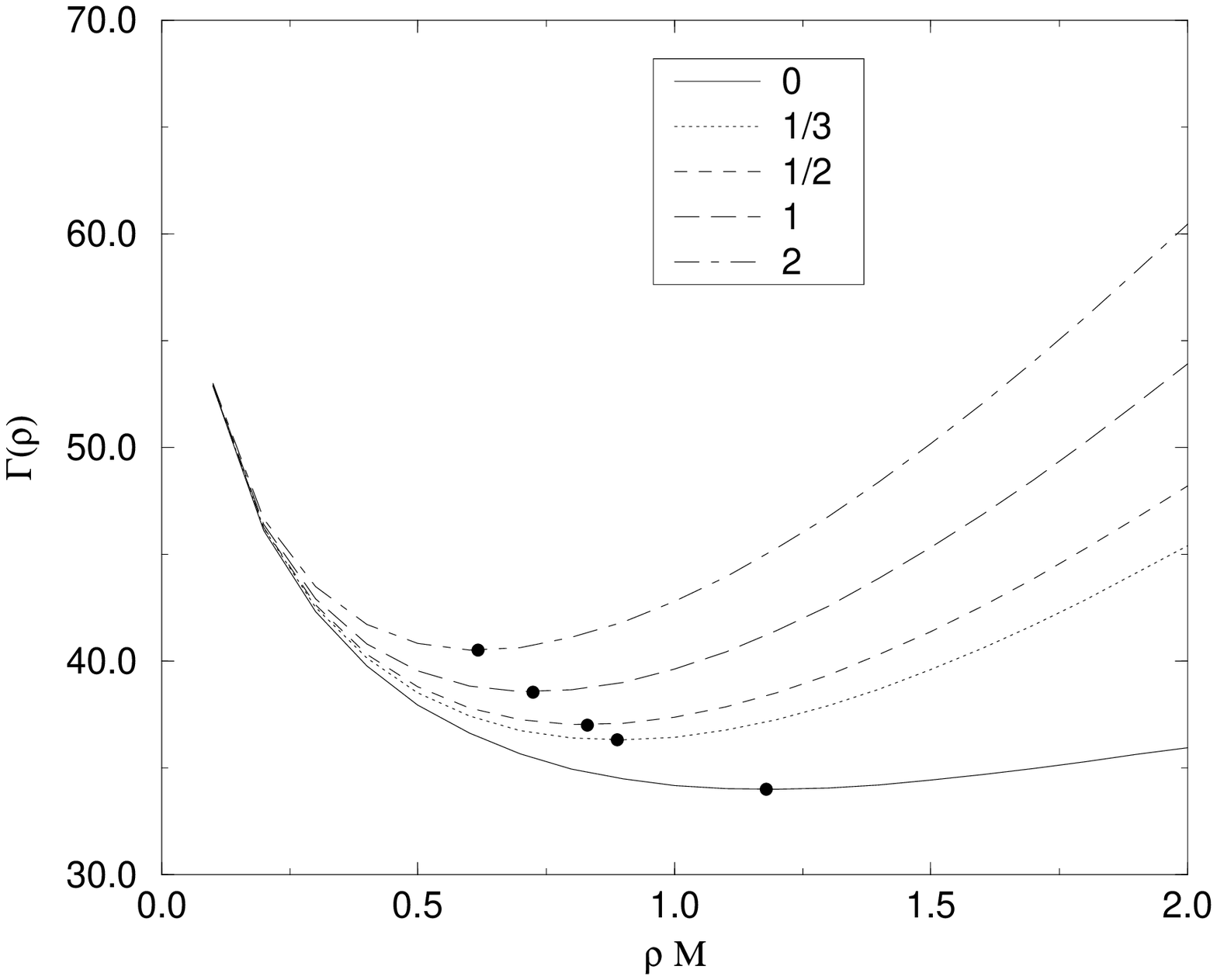}
\caption{The classical action plus one-loop logarithmic correction for
the profile $f_2$ as a function of the instanton size. The different
curves correspond to different values of the parameter $a = 0, \ldots,
2$, and the bullets mark the minimum of each curve.} \label{fig3}
\end{center}
\end{figure}

\begin{figure}
\begin{center}
\leavevmode
\epsfxsize=4.3in
\epsfbox{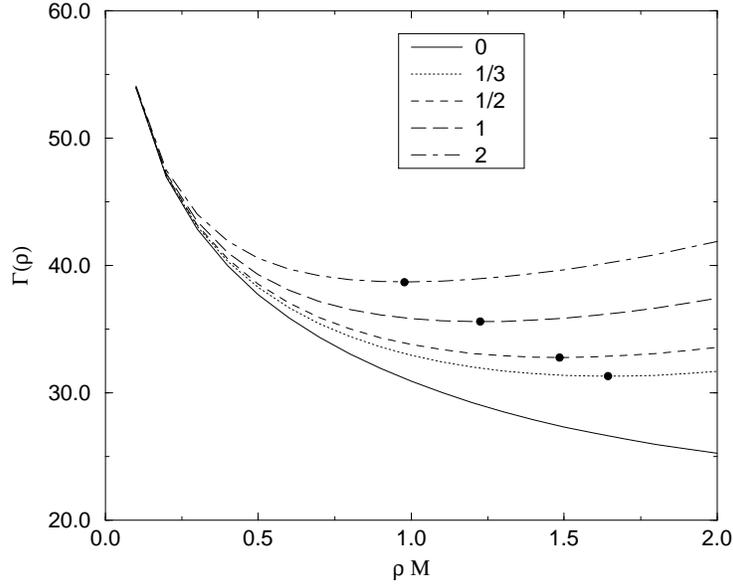}
\caption{The classical action plus one-loop logarithmic correction for
the profile $f_{\rm AM}$ as a function of the instanton size. The different
curves correspond to different values of the parameter $a = 0, \ldots,
2$, and the bullets mark the minimum of each curve.} \label{fig4}
\end{center}
\end{figure}

In all cases the action has
a minimum at a size of the order of $1/M$.  The values for the size of
the instanton, its action and the curvature at the minimum are given in
Table \ref{tab} for different values of $a$.
The results for the Atiyah-Manton Ansatz and our function $f_1$ are
practically indistinguishable. Varying the parameter $a$ between $1/2$
and $2$ the size of the stable instanton varies between $1.5/M$ and
$1/M$.
The Ansatz $f_2$ leads to the instanton size smaller by a factor of
$1.5-2$.
We consider however the former estimate to be better, since the Ansatz
$f_2$ gives a considerably larger value of the action (see Fig. 2)
and is therefore
not a very good choice for the instanton ``valley'' configuration.

\begin{table}
\begin{tabular}{cddddddddd}
 & \multicolumn{3}{c}{$f_1$} & 
  \multicolumn{3}{c}{$f_2$} & 
  \multicolumn{3}{c}{$f_{\rm AM}$} \\
\cline{2-4}
\cline{5-7}
\cline{8-10}
$a$ & 
	$\rho_{\rm min}$ & $\Gamma(\rho_{\rm min})$ & 
	$\Gamma''(\rho_{\rm min})$ &
 	$\rho_{\rm min}$ & $\Gamma(\rho_{\rm min})$ & 
	$\Gamma''(\rho_{\rm min})$ &
 	$\rho_{\rm min}$ & $\Gamma(\rho_{\rm min})$ & 
	$\Gamma''(\rho_{\rm min})$ \\
\hline
1/3  & 1.5 & 32. & 6.9 & 0.9 & 36. & 23. & 1.6 & 32. & 6.3 \\
1/2  & 1.4 & 33. & 7.9 & 0.8 & 37. & 26. & 1.5 & 33. & 7.0 \\
1  & 1.1 & 36. & 9.7 & 0.7 & 39. & 33. & 1.2 & 36. & 8.1 \\
2 & 0.9 & 39. & 12.0 & 0.6 & 41. & 41. & 1.0 & 39. & 9.7 \\
\end{tabular}
\caption{Stable size $\rho_{\rm min}$ for the soliton in units of $1/M
= 0.75 GeV^{-1}$, the value of the minimum action $\Gamma(\rho_{\rm min})$
and the curvature at the minimum $\Gamma''(\rho_{\rm min})$, for the three
profiles, for different values of the parameter $a$.}
	\label{tab}
\end{table}

The numerical analysis of this section confirms therefore our
expectations. The presence of the dynamical mass stabilizes the size
of the instantons at the value $\rho=\frac{1-1.5}{M}$. The large
instanton problem therefore finds a nonperturbative solution in the
framework of the Gaussian variational approximation.

\section{Discussion}\label{sec:discussion}

In this paper we have discussed how the instanton transitions appear
in the framework of the gauge invariant Gaussian approximation of
\cite{KK95}.
The relative gauge rotation between two gauge field configurations
between which the tunneling transition is most probable is determined
from the saddle point equation of the effective nonlinear $\sigma$-model.
We found that the relative gauge transformations for most probable
tunneling transitions are very similar to the ones found in the
standard path
integral instanton approach.

The value of the logarithm of the 
tunneling probability for small size instantons
however turns out to be larger by about a factor of two in the
variational vacuum. This is understandable since the tail of a Gaussian wave
function decreases faster and therefore the overlap integral
of two such wave functions is smaller than of the semi-classical wave
functions.

Our main result concerns the effect of the dynamical mass scale that
characterizes the variational vacuum. We find that the presence of
this scale stabilizes the size of the instanton. The instantons of the
size larger than the inverse of this scale are suppressed. When
account is taken of the proper running of the strong coupling
constant,
the integral over the instanton sizes has a saddle
point. This saddle point determines the most likely size of the instanton 
as $\rho=1-1.5 / M$. 
The large size instanton infrared problem which plagues the standard
dilute instanton gas approximation is thereby removed in the
variational vacuum due to the presence of the dynamical
nonperturbative scale.

An interesting point is that the instanton action for the most likely size
instanton is pretty large, the numerical value being around 35 (see
Table \ref{tab}). This means that the configurations with small number of
instantons and anti-instantons are not important energetically. 
Nevertheless, the generation of the dynamical mass itself within the
framework of our variational approximation is due to the instanton
condensation. This is so since the effective $\sigma$-model is in the
disordered phase at the value of the best variational parameter $M$.
This suggests that the important type of configurations are the ones
that contain many instantons and anti-instantons. The small fugacity
factor in these configurations can be overcome by a large entropy and
also by the interaction between instantons and anti-instantons. This
interaction is known to be attractive for some relative color
orientation. The interaction also is quite long range, since
the instanton profile function away from the instanton center
decreases only as the second power of the distance.
It is therefore entirely possible that the important configurations
for the energy minimization
are of the type of the instanton liquid--namely those having large
number of instantons which although fairly dilute nevertheless feel
each others presence strongly due to the long range interaction.
It is in fact interesting to note that the most likely instanton size
we found here is consistent with the average size of the instantons in
the instanton liquid model of \cite{S82,DP84}.
For the case of $SU(2)$, the average instanton size, in units of the gluon
condensate obtained in the instanton liquid model, turns out to be 
\cite{DP84},
\[
\rho \, 
	\left( 
	\langle F_{\mu \nu}^a F_{\mu \nu}^a \rangle
	\alpha/\pi \right)^{1/4}
	\sim .4 .
\]
In our case, taking the value of the gluon condensate obtained in the
variational approach \cite{KK95}, we find,
\[
\rho \, \left(
	\langle F_{\mu \nu}^a F_{\mu \nu}^a \rangle \alpha/\pi 
	\right)^{1/4}
	\sim .2 - .3 .
\]

The relation of the variational approach with the instanton liquid
model is a very interesting open question and warrants further study.

\acknowledgements

The results in this paper were presented in June 1998 at the NORDITA
workshop ``Instantons and monopoles in the QCD vacuum''.  I.I.K. is
grateful to the organizers of the workshop for hospitality and to D.I.
Diakonov, V.I. Petrov, Yu.A. Simonov and K. Zarembo for interesting
discussions.  W.E.B. wishes to thank PPARC for a research studentship.
The work of J.P.G. was supported by EC Grant ARG/B7-3011/94/27. A.K. is
supported by a PPARC advanced fellowship.

\appendix
\section*{}

The finite action for the case of the mass scale $M = 0$ is given in
(\ref{acfin}). To obtain the explicit action for the profile of the
hedgehog field (\ref{skyrmion}) it is necessary to calculate the
Laplacian over $v = x - y$ of the product of the right currents
$\lambda^a_i$ at the points $x$ and $y$, which has the form
\beq
\lambda_i^a(x) \lambda_i^a(y) = 
	\frac{1}{g^2} \sum_{n=0}^2 \cos^n \theta H_n(r,s) .
\eeq
The expression for $H_n$ are obtained from (\ref{lili}),
\begin{eqnarray}
H_0 &=& \frac{1}{r s} \sin{2 f(r)} \sin{2 f(s)} +
	\frac{2}{r} \sin{2 f(r)} f'(s) +
	\frac{2}{s} \sin{2 f(s)} f'(r) , \\
H_1 &=& \frac{8}{r s} \sin^2 f(r) \sin^2 f(s) , \\
H_2 &=& \frac{1}{r s} \left( \sin{2 f(r)} - 2 r f'(r) \right)
	\left( \sin{2 f(s)} - 2 s f'(s) \right) .
\end{eqnarray}
After calculating the Laplacian over $v$ we obtain
\beq
\left(\frac{\partial^2}{\partial x_j^2} 
	+ \frac{\partial^2}{\partial y_j^2} - 
	2 \frac{\partial^2}{\partial x_j \partial y_j} \right) 
	\left[\lambda_i^a(x) \lambda_i^a(y) \right] = \frac{1}{g^2} 
	\sum_{n=0}^3 \cos^n \theta \tilde{H}_n(r,s) .
\eeq
From Eq. (\ref{lili}) the expressions for $\tilde{H}_n(r,s)$ can
be found:
\begin{eqnarray}
\lefteqn{2 r^3 s^3 \tilde{H}_0(r,s) =  
	8\,r\,s\, \sin^2 f(r) \, \sin^2 f(s)  + 
	{r^2}\, \sin{2\,f(r)} \, \sin{2\,f(s)} }  
	\nonumber \\ &&
	- 16\,{r^2}\,s\, \cos{f(r)} \,
	 \sin{f(r)} \, \sin^2 f(s) \,f'(r)  
	- 2\,{r^3}\, \sin{2\,f(s)} \,f'(r) 
	\nonumber \\ &&
	- 2\,{r^2}\,{s^2}\, \sin{2\,f(r)} \, \sin{2\,f(s)} \,{{f'(r)}^2}  
	- 2\,{r^2}\,s\, \sin{2\,f(r)} \,f'(s)
	\nonumber \\ &&
	+ 4\,{r^3}\,s\,f'(r)\,f'(s)  
	- 4\,{r^3}\,{s^2}\, \sin{2\,f(s)} \,f'(r)\,{{f'(s)}^2}  
	+ 2\,{r^2}\,{s^2}\, \sin{2\,f(s)} \,f''(r) 
	\nonumber \\ &&
	+ {r^2}\,{s^2}\, \cos{2\,f(r)} \, \sin{2\,f(s)} \,f''(r)  
	+ 2\,{r^3}\,{s^2}\, \cos{2\,f(s)} \,f'(r)\,f''(s)  
	\nonumber \\ &&
	+ {r^3}\,{s^2}\, \sin{2\,f(s)} \,f^{(3)}(r) + 
	(r \leftrightarrow s) ,
	\\
\lefteqn{2 r^3 s^3 \tilde{H}_1(r,s) =   
	-8\,{r^2}\, \sin^2 f(r) \, \sin^2 f(s)   
	+ 3\,r\,s\, \sin{2\,f(r)} \, \sin{2\,f(s)} } 
 	\nonumber \\ &&
	- 4\,{r^2}\,s\, \sin{2\,f(s)} \,f'(r)  
	- 2\,{r^2}\,s\, \cos{2\,f(r)} \, \sin{2\,f(s)} \,f'(r)  
	\nonumber \\ &&
	+ 8\,{r^2}\,{s^2}\,{{ \cos{f(r)} }^2}\,
	 \sin^2 f(s) \,{{f'(r)}^2}  
	+ 8\,{r^2}\,{s^2}\, \cos{2\,f(r)} \,f'(r)\,f'(s)  
	\nonumber \\ &&
	- 4\,{r^2}\,{s^2}\, \cos{2\,f(r)} \,
	 \cos{2\,f(s)} \,f'(r)\,f'(s)  
	- 8\,{r^2}\,{s^2}\, \sin^2 f(r) \,
	 \sin^2 f(s) \,{{f'(s)}^2}  
	\nonumber \\ &&
	+ 8\,{r^2}\,{s^2}\, \cos{f(r)} \,
	 \sin{f(r)} \, \sin^2 f(s) \,f''(r)  
	+ 6\,{r^3}\,s\, \sin{2\,f(s)} \,f''(r) 
	\nonumber \\ &&
	- 8\,{r^3}\,{s^2}\,f'(s)\,f''(r)  
	- 4\,{r^3}\,{s^2}\, \cos{2\,f(s)} \,f'(s)\,f''(r)  + 
	(r \leftrightarrow s) , \\
\lefteqn{2 r^3 s^3 \tilde{H}_2(r,s) =  
	-32\,r\,s\, \sin^2 f(r) \, \sin^2 f(s)  - 
	3\,{r^2}\, \sin{2\,f(r)} \, \sin{2\,f(s)}  } 
 	\nonumber \\ &&
	+ 32\,{r^2}\,s\, \cos{f(r)} \,
	 \sin{f(r)} \, \sin^2 f(s) \,f'(r)  
	+ 6\,{r^3}\, \sin{2\,f(s)} \,f'(r) 
 	\nonumber \\ &&
	- 2\,{r^2}\,{s^2}\, \sin{2\,f(r)} \,
	 \sin{2\,f(s)} \,{{f'(r)}^2}  
	+ 6\,{r^2}\,s\, \sin{2\,f(r)} \,f'(s) 
	\nonumber \\ &&
	- 12\,{r^3}\,s\,f'(r)\,f'(s)  
	- 32\,{r^2}\,{s^2}\, \cos{f(r)} \,
	 \cos{f(s)} \, \sin{f(r)} \, \sin{f(s)} \, f'(r)\,f'(s) 
	\nonumber \\ &&
	+ 4\,{r^2}\,{s^3}\, \sin{2\,f(r)} \,{{f'(r)}^2}\,f'(s)   
	- 2\,{r^2}\,{s^2}\, \sin{2\,f(s)} \,f''(r) 
	\nonumber \\ &&
	+ {r^2}\,{s^2}\, \cos{2\,f(r)} \, \sin{2\,f(s)} \,f''(r) + 
	4\,{r^2}\,{s^3}\,f'(s)\,f''(r)  
	\nonumber \\ &&
	- 2\,{r^3}\,{s^2}\, \cos{2\,f(s)} \,f'(r)\,f''(s) - 
	{r^3}\,{s^2}\, \sin{2\,f(s)} \,f^{(3)}(r) 
	\nonumber \\ &&
	+ 2\,{r^3}\,{s^3}\,f'(s)\,f^{(3)}(r) + 
	(r \leftrightarrow s) , \\
\lefteqn{2 r^3 s^3 \tilde{H}_3(r,s) = 
	-9\,r\,s\, \sin{2\,f(r)} \, \sin{2\,f(s)}  + 
	12\,{r^2}\,s\, \sin{2\,f(s)} \,f'(r) }
	\nonumber \\ &&
	+ 6\,{r^2}\,s\, \cos{2\,f(r)} \, \sin{2\,f(s)} \,f'(r) 
	- 16\,{r^2}\,{s^2}\,f'(r)\,f'(s) 
	\nonumber \\ &&
	- 8\,{r^2}\,{s^2}\, \cos{2\,f(r)} \,f'(r)\,f'(s)
	- 4\,{r^2}\,{s^2}\, \cos{2\,f(r)} \,
	 \cos{2\,f(s)} \,f'(r)\,f'(s) 
	\nonumber \\ &&
	- 6\,{r^3}\,s\, \sin{2\,f(s)} \,f''(r) + 
	8\,{r^3}\,{s^2}\,f'(s)\,f''(r)
	+ 4\,{r^3}\,{s^2}\, \cos{2\,f(s)} \,f'(s)\,f''(r)
	\nonumber \\ &&
	- 4\,{r^3}\,{s^3}\,f''(r)\,f''(s) + 
	(r \leftrightarrow s) .
\end{eqnarray}

The expressions for the angular integrals (\ref{int}) appearing in the
action (\ref{action0}) are the following:
\begin{eqnarray}
I_0 &=& \frac{1}{2rs} \log \left[ \frac{(r+s)^2}{(r-s)^2} \right] , 
	\\ \nonumber
I_1 &=& - \frac{1}{rs} + \frac{(r^2 + s^2)}{4r^2 s^2} \log 
	\left[ \frac{(r+s)^2}{(r-s)^2} \right] , 
	\\ \nonumber
I_2 &=& -\frac{1}{2r^2} - \frac{1}{2s^2} + 
	\frac{(r^2+s^2)^2}{8 r^3 s^3} \log 
	\left[ \frac{(r+s)^2}{(r-s)^2} \right] , 
	\\ \nonumber
I_3 &=& -\frac{1}{3rs} - \frac{(r^2 + s^2)^2}{4r^3s^3} +
	\frac{(r^2 + s^2)^3}{16 r^4 s^4} 
	\log \left[ \frac{(r+s)^2}{(r-s)^2} \right] .
\end{eqnarray}
The corresponding integrals (\ref{intm}) for the case of $M \neq 0$ read:
\begin{eqnarray}
\tilde{I}_0 &=& \frac{1}{rs} 
	\left( \frac{\sin^2(t/2)}{t^2} \right)_{M |r-s|}^{M (r+s)} ,
	\\
\tilde{I}_1 &=& \frac{r^2 + s^2}{2 r s} \tilde{I}_0 + 
	\frac{1}{2 r^2 s^2} 
	\left( \frac{\cos{t}}{2} - {\rm ci}(t) + 
	\ln(t/2) \right)_{M |r-s|}^{M (r+s)} ,
	\\
\tilde{I}_2 &=& \frac{r^2 + s^2}{2rs} \tilde{I}_1 - 
	\frac{1}{8r^3s^3} \left[ (-4 + t^2 - r^2 - s^2) \cos{t}
	\right. 
	\\ \nonumber
	& & + \left. 2 (r^2 + s^2) 
	({\rm ci}(t) - \ln{(t/2)}) + 
	t(t - 4 \sin{t}) \right]_{M |r-s|}^{M (r+s)} ,
\end{eqnarray}
where ${\rm ci}(t)$ is the Cosine-integral function
\begin{equation}
{\rm ci}(t) = - \int_t^{\infty} \frac{\cos{t}}{t} dt .
\end{equation}

\end{document}